\documentstyle[12pt]{article}
\begin{document}
\begin{flushright}
hep-th/9707004
\end{flushright}
\vskip .7cm
\begin{center}
{\bf {DYNAMICS IN A NONCOMMUTATIVE PHASE SPACE}}

\vskip 2cm

{\bf R.P.Malik,\footnote{On leave of absence from Bogoliubov Laboratory
 of Theoretical Physics, JINR, Dubna, Moscow, Russia.}
\footnote{ e-mail address: malik@thsun1.jinr.dubna.su}
A. K. Mishra
\footnote {e-mail address: mishra@imsc.ernet.in}
and G. Rajasekaran}
\footnote{ e-mail address: graj@imsc.ernet.in}\\
{\it Institute of Mathematical Sciences} \\
{\it C.I.T. Campus, Madras - 600 113, India} \\

\vskip 2.5cm

{\bf ABSTRACT}
\end{center}

Dynamics has been generalized to a noncommutative phase space.
The noncommuting phase space is taken to be invariant under
the quantum group $GL_{q,p}(2)$. The $q$-deformed differential
calculus on the phase space is formulated and using this,
both the Hamiltonian and
Lagrangian forms of dynamics have been constructed.
In contrast to earlier forms of $q$-dynamics, our formalism
has the advantage of preserving the conventional symmetries
such as rotational or Lorentz invariance.

\baselineskip=16pt

\vskip 1cm
%\noindent PACS Nos : 03.70; 05.30; 02.20; 71.27

\noindent
{\bf 1 Introduction}\\

\noindent
Quantum groups, $q$-deformations and the associated noncommutative 
quantum plane have been a subject of considerable interest in
mathematical physics during the past few years [1--17]. 
However, despite many attempts [18--25], the key concepts of quantum 
groups have
 not yet found a firm footing in the domain of physical applications.\\

Recently Lukin, Stern and Yakushin 
[19]  developed a dynamical formalism on a 2D quantum plane 
by exploiting the notion of a 
tangent space defined over a 2D $q$-deformed configuration space.
In this approach, a  differential calculus, that is 
invariant under  the quantum group $ GL_{q,p}(2) $,
is developed and this is then applied to
the construction of  dynamics. The conventional symmetries such as
 rotational invariance and Lorentz invariance are 
however lost in this dynamics. 
Furthermore, the status of a single dimensional system is not clear.
In an altogether different approach, Aref'eva and Volovich [18] 
introduced the
deformation parameter in the phase space for the description of the free
nonrelativistic particle and the  harmonic oscillator
which was then  extended to the multidimensional case [20]. However, 
a consistent differential calculus in the phase space was not 
developed in these works, and as a consequence, some of their results
are of an adhoc nature.\\ 

In the present work, we shall take the phase space as a 2D quantum plane
based on the quantum group
$GL_{q,p}(2)$ and try to construct  dynamics of one-dimensional
as well as multidimensional systems, based on a consistent differential
calculus. We shall adapt the 2D differential calculus of Lukin etal.
for our purpose. We are thus able to complete the program initiated
in Refs. [18, 20] by using the mathematical tools invented by 
Lukin etal. [19]. The resulting dynamics, although $q$-deformed,
preserves conventional symmetries such as rotational and
Lorentz invariance.\\

The material of  our work is organised as follows. In Sec. 2, we develop
a $GL_{q,p}(2)$ invariant differential calculus in the 2D phase space of
a one dimensional system and construct dynamics 
on the $q$-deformed symplectic manifold. We also furnish examples of dynamics
for a general class of Lagrangians  
expressed  as a polynomial of second degree in  position and velocity. We
develop a differential calculus in the multidimensional phase space 
in Sec. 3 and show  that the mutual consistency of the $GL_{q,p}(2)$ 
invariance of the phase space with the separate $GL(N)$ invariance
of the configuration and momentum spaces requires the parameters to satisfy
$ pq = 1 $. We then construct the  multidimensional dynamics and discuss
 examples including a free massless relativistic
particle. Finally, Sec. 4
is devoted to summary and discussion.\\

\noindent             
{\bf 2 Single dimensional systems}\\

\noindent
{\it  2.1  Differential calculus}\\

\noindent
We construct here a $GL_{q,p}(2)$-invariant differential calculus for
a dynamical system
on a $2D$ cotangent manifold (i.e., phase space) corresponding
to a one-dimensional
configuration manifold. 
The phase space variables, the coordinate $x$ and momentum $\pi$,
are functions of a commuting evolution parameter $t$ in terms of which the
trajectory of the physical system is parametrized in the phase space.
We introduce a variational derivative $\delta$ which is identified with
the exterior derivative of the differential geometry
$d$ (i.e., $ d^2 = \delta^2 = 0$). Our starting point
is the following set of basic 
relations among $ x, \pi, \delta x $ and $ \delta \pi $:
$$
\begin{array}{lcl}
x\;\pi &=& q \;\pi\;x, \nonumber\\
x\;\delta x &=& pq \;\delta x\; x , \nonumber\\
x\;\delta \pi &=& q\; \delta \pi \; x + (pq -1)\delta x\; \pi,
\nonumber\\
\pi\;\delta  \pi &=& pq \;\delta  \pi\; \pi, \nonumber\\
\pi\; \delta x &=& p\;\delta x \;\pi,
\end{array} \eqno(2.1)
$$
where $q,p $ are nonzero c-numbers which may be complex.\\

The algebra given by (2.1) is  invariant under the following $GL_{q,p}(2)$ 
transformation 
$$
\begin{array}{lcl}
\left (\begin{array}{c}
\phi \\
\psi \\   \end{array} \right) \;
\rightarrow
\left (\begin{array}{cc}
A  & B \\
C  & D \\   \end{array} \right) \;
\left (\begin{array}{c}
\phi \\
\psi \\   \end{array} \right) ,
\end{array} \eqno(2.2)
$$
where the pair ($ \phi, \psi $ ) stands for ($ x, \pi $) or
($ \delta x, \delta \pi $) which commutes with the elements $ A,B,C $ and
$D$ of a $ 2 \times  2 $ $GL_{q,p}(2)$ matrix obeying the braiding relations
in rows and columns as given below
$$
\begin{array}{lcl}
A\;B &=& p\; B\;A, \quad A\;C = q\; C\;A,
\quad B\;C = (q/p)\;C\;B, \quad B\;D = q\;D\;B, \nonumber\\
C\;D &=&p\;D\;C, \quad AD-DA = (p - q^{-1}) BC = (q - p^{-1}) CB.\\
\end{array} \eqno(2.3)
$$

We shall use the definition of the partial derivative 
$$
\begin{array}{lcl}
\delta F = \delta x \;\Bigl ( {\displaystyle \frac {\partial F}
{\partial x}} \Bigr )\; +\; \delta \pi\;
\Bigl ( {\displaystyle \frac{\partial F} {\partial \pi}} \Bigr ),
\end{array} \eqno(2.4)
$$
and the Leibnitz rule [11]
$$
\begin{array}{lcl}
\delta ( F G  ) = \;(\delta F)\; G \;+\; F \; (\delta G),
\end{array} \eqno(2.5)
$$
where $F$ and $G$ are polynomial functions of $x$ and $\pi$.\\

We next define $ \delta x = \dot x \;\delta t$ and
 $\delta \pi = \dot \pi\; \delta t$ and get from (2.1), the 
relations 
$$
\begin{array}{lcl}
x\;\dot x &=& pq \;\dot x\; x , \nonumber\\
x\;\dot \pi &=& q\; \dot \pi \; x + (pq -1)\dot x\; \pi,
\nonumber\\
\pi\;\dot  \pi &=& pq \;\dot  \pi\; \pi, \nonumber\\
\pi\; \dot x &=& p\;\dot x \;\pi.
\end{array} \eqno(2.6)
$$
Following Ref. [19], we also postulate
$$
\begin{array}{lcl}
\dot x\; \dot \pi = q\; \dot \pi\; \dot x.
%\label{2.5}
\end{array}\eqno(2.7)
$$  
All $qp$-relations with a single variation and single ``time'' derivative 
can be derived from (2.1) and (2.6) by exploiting Leibnitz rule (2.5). 
These are listed below 
$$
\begin{array}{lcl}
\dot x\;\delta  x &=&  \delta  x\; \dot x,\nonumber\\
\dot \pi\;\delta  \pi &=&  \delta  \pi\; \dot \pi, \nonumber\\
\dot x\;\delta  \pi &=& q\; \delta  \pi\; \dot x,\nonumber\\
x\;\delta \dot x &=& pq \;\delta \dot x\; x + (pq -1)\;\delta x\; \dot x,
\nonumber\\
\pi\;\delta \dot \pi &=& pq \;\delta \dot \pi\; \pi
+ (pq -1)\;\delta \pi\; \dot \pi,\nonumber\\
x\;\delta \dot \pi &=& q \;\delta \dot \pi\; x
+ (pq -1)\;(\delta \dot x \;\pi + \delta x\;\dot \pi ),\nonumber\\
\pi\;\delta \dot x &=& p \;\delta \dot x\; \pi
+ (pq -1)\;\delta \pi\; \dot x,\nonumber\\
\dot \pi\;\delta  x &=& p \;\delta x\; \dot \pi
- (pq -1)\;\delta \pi\; \dot x.
\end{array}\eqno(2.8)
$$
The relations involving ``time'' derivatives and their variations are
$$
\begin{array}{lcl}
\dot x\;\delta \dot x &=& qp\; \delta \dot x\; \dot x,\nonumber\\
\dot \pi\;\delta \dot \pi &=& qp\; \delta \dot \pi\; \dot \pi, \nonumber\\
\dot x\;\delta \dot \pi &=& q \;\delta \dot \pi\; \dot x
+ (pq -1)\;\delta \dot x \;\dot \pi ,\nonumber\\
\dot \pi\;\delta \dot x &=& p \;\delta \dot x\; \dot \pi .
\end{array}\eqno(2.9)
$$
The following wedge products complete the full algebraic
structure of the differential calculus
$$
\begin{array}{lcl}
\delta x\;\wedge\;\delta \dot x
&=& -\; \delta \dot x\; \wedge\;\delta x,\nonumber\\
\delta \pi\;\wedge\;\delta \dot \pi
&=& -\; \delta \dot \pi\;\wedge\; \delta \pi, \nonumber\\
\delta x\;\wedge\;\delta \pi
&=& - p^{-1}\; \delta \pi\; \wedge\;\delta x,\nonumber\\
\delta \dot x\;\wedge\;\delta \dot \pi
&=& -p^{-1}\; \delta \dot \pi\; \wedge\;\delta \dot x,\nonumber\\
\delta \pi\;\wedge\;\delta \dot x
&=& -\; q^{-1}\;\delta \dot x\;\wedge\; \delta \pi, \nonumber\\
\delta x\;\wedge\;\delta \dot \pi
&=& -\; p^{-1}\;\delta \dot \pi\; \wedge\;\delta x
+ (pq)^{-1}\; (1 - pq)\;\delta \dot x \;\wedge\;\delta \pi.
\end{array}\eqno(2.10)\\
$$

All the $qp$-algebraic relations (2.1) , (2.6--2.10) can be shown to
be invariant under the $GL_{q,p}(2)$ transformations (2.2) if  the pair
($ \phi,\psi $) are taken to be ($ x, \pi$), ($\delta x, \delta \pi$),
($\dot x, \dot \pi$) or ($ \delta \dot x, \delta \dot \pi$). 
Furthermore, it can be checked that when
any of these quantities $\phi$ or $ \psi$ is
commutated through the above algebraic relations, no new secondary relations
emerge. This establishes the associativity conditions which are equivalent
to the validity of the Yang-Baxter equations.\\

The construction of the algebraic structure of the $qp$-deformed
differential calculus given in (2.1) and (2.6--2.10) parallels the
work of Lukin etal. [19]. Whereas  Lukin etal. apply it to the two
dimensional configuration space, we apply it to the two dimensional
phase space.\\

\noindent
{\it  2.2  Dynamics} \\
                    
\noindent
We begin with the action $S$ which is expressed in terms of the   
Hamiltonian function $ H(x,\pi) $ as
$$
\begin{array}{lcl}
S= \int dt\; \bigl [\; \pi\;\dot x - H (x,\pi) \;\bigr ].
%\label{2.9}
\end{array}\eqno(2.11)
$$
 Using (2.4) and (2.5), the action principle can be written as
$$
\begin{array}{lcl}
\delta S = 0 = \int dt \Bigl [ \;\; \delta \pi\; \dot x
+ \pi\; \delta \dot x -  \delta x\; {\displaystyle
\frac {\partial H} {\partial x}} 
-  \delta \pi\; {\displaystyle \frac{\partial H} {\partial \pi}}\;\;
\Bigr ].
\end{array}\eqno(2.12)
$$
The following Hamilton equations of motion emerge
$$
\begin{array}{lcl}
\dot x = \frac{\displaystyle 1}{\displaystyle pq} 
\frac{\displaystyle \partial H}
{\displaystyle \partial \pi}, 
\end{array}
$$

$$
\begin{array}{lcl}
\dot \pi = -\;\frac{\displaystyle 1}{\displaystyle p} 
\frac{\displaystyle \partial H}
{\displaystyle \partial x}, 
\end{array} \eqno(2.13)
$$
when we use
$$
\pi\;\delta \dot x = p\; \delta \dot x\;\pi
+ (pq - 1) \delta  \pi\;\dot x,
$$
{}from (2.8) while taking all the variations to the left. If we now
define the $qp$-deformed Poisson bracket between two dynamical variables
$F(x,\pi) $ and $G (x,\pi)$ as
$$
\begin{array}{lcl}
\{ F, G \} = \frac{\displaystyle 1}{\displaystyle qp} 
\frac{\displaystyle \partial G}{\displaystyle \partial \pi} 
\frac{\displaystyle \partial F}{\displaystyle \partial x} - 
\frac{\displaystyle 1}{\displaystyle p} 
\frac{\displaystyle \partial G}{\displaystyle \partial x} 
\frac{\displaystyle \partial F}{\displaystyle \partial \pi},
%\label{2.11}
\end{array} \eqno(2.14)
$$
the equations of motion (2.13) become
$$
\begin{array}{lcl}
\dot x & = & \{ x, H \},  \;\quad  \nonumber\\ 
\dot \pi & = & \{ \pi, H \}, 
%\label{?}
\end{array} \eqno(2.15)
$$
and the basic canonical brackets turn out to be
$$
\begin{array}{lcl} 
\{ x, \pi \} &=& (pq)^{-1}, \;  \nonumber\\ 
\{ \pi, x \} &=& -\;p^{-1}, \;  \nonumber\\
\{ x, x   \} &=& \{ \pi, \pi \} = 0. 
\end{array}\eqno(2.16)
$$
It may be noted that the above basic brackets remain invariant under the
$SL_{q}(2)$ transformations defined by (2.2) and (2.3) with  $q=p$ and
the corresponding $q$-determinant is chosen to be unity, i.e.;
$$
\begin{array}{lcl}
A\;D - q\;B\;C = D\;A - q^{-1} C\;B  = 1.\\
\end{array}
$$

We now turn to the Lagrangian dynamics. We obtain the Lagrangian function
$ L(x,\dot x) $ by the Legendre transformation
$$
\begin{array}{lcl} 
L = \pi\;\dot x - H(x,\pi),
\end{array} 
$$
where, on the r.h.s., the canonical momentum ($\pi$) is to be eliminated
using
$$
\begin{array}{lcl}
\pi    =   \frac{\displaystyle 1}{\displaystyle p}\; 
\frac{\displaystyle \partial L}
{\displaystyle \partial \dot x}.
\end{array}\eqno(2.17)
$$
Rewriting the action $S= \int dt L(x,\dot x)$ and requiring \\ 
$$
\begin{array}{lcl} 
\delta S = 0    \equiv \int dt\; \bigl (\; \delta x\; 
 \frac{\partial L}{\partial  x} + \; \delta \dot x\;
 \frac{\partial L}{\partial \dot x}\; \bigr ),  
%\label{2.15}
\end{array}\eqno(2.18)
$$
we obtain the Euler-Lagrange equations of motion\\
$$
\begin{array}{lcl}
\frac{\displaystyle \partial L}
{\displaystyle \partial  x} = \;
\frac{\displaystyle d}{\displaystyle dt} 
\Bigl ( 
 \frac{\displaystyle \partial L}
{\displaystyle \partial \dot x} \Bigr ),
\end{array}\eqno(2.19)
$$
which is consistent with the Hamilton equations (2.13) if 
we take into account the definition of momentum
 (2.17).\\

An important feature of the $q$-deformed dynamics is the existence of
certain restrictions on the form of the Lagrangian. These follow from the  
$qp$-commutation relations: 
$$
\begin{array}{lcl}
x\;\pi &=& q\; \pi\;x, 
\quad \dot x\; \dot \pi = q\; \dot \pi\; \dot x,
\quad \pi\;\dot \pi= pq\; \dot \pi\;\pi\;,\nonumber\\
\pi\;\dot x &=& p\;\dot x\;\pi, \quad
x\;\;\dot \pi = q\; \dot \pi\; x + (pq -1)\;\dot x\;\pi,\quad
\end{array}
$$
and these restrictions are
$$
\begin{array}{lcl}
x\;\frac{\displaystyle \partial L}{\displaystyle \partial \dot x}
 =  q\; \frac{\displaystyle \partial L}{\displaystyle \partial \dot x}
\;x,\qquad \;\dot x\;\frac{\displaystyle \partial L}
{\displaystyle \partial  x} = q\;
\frac{\displaystyle \partial L}{\displaystyle \partial  x}\;\dot x,
\end{array}
$$
$$
\begin{array}{lcl}
\frac{\displaystyle \partial L}{\displaystyle \partial \dot 
x}\;
\frac{\displaystyle \partial L}{\displaystyle \partial  x}  =  pq\;
\frac{\displaystyle \partial L}{\displaystyle \partial  x}
\frac{\displaystyle \partial L}{\displaystyle \partial \dot x}, \qquad
\;\frac{\displaystyle \partial L}
{\displaystyle \partial \dot x}\;\dot x = p\;\dot x\;
\frac{\displaystyle \partial L}{\displaystyle \partial \dot x}, \;
\end{array}
$$
$$
\begin{array}{lcl}
x\;\frac{\displaystyle \partial L}{\displaystyle \partial  x} 
=  q\;
\frac{\displaystyle \partial L}{\displaystyle \partial  x}\; x
+ \;(pq -1)\;\dot x\;\frac{\displaystyle \partial L}
{\displaystyle \partial \dot x}.
\end{array}\eqno(2.20)
$$
Similar restrictions can be written for the Hamiltonian  too.\\

More generally we may start with the following action  defined 
on a $qp$-deformed symplectic manifold
$$
\begin{array}{lcl}
S = \int dt \Bigl [\;\; {\displaystyle \frac{1} {1 + pq}}\;\; 
{\displaystyle \sum_{M, N}} \Omega_{MN}\;z^{M}\;\dot z^{N}
-\;   H(z) \;\;\Bigr ] , 
\end{array}\eqno(2.21)
$$
where $ z^M = (x, \pi)$ for $ M = 1,2 $ is an arbitrary 
phase space coordinate
on the symplectic manifold and $\Omega_{MN}$ is a covariant symplectic
metric that satisfies
$$
\begin{array}{lcl}
 {\displaystyle \sum_{N}} \Omega_{MN} \Omega^{NL}  =  \delta_{M}^{L}
= 
{\displaystyle \sum_{N}} \Omega^{LN} \Omega_{NM}, 
\end{array}
$$
$$
\begin{array}{lcl}
\Omega_{MN}  = \; - q\; \Omega_{NM}, \qquad   M \geq N.
\end{array}\eqno(2.22)
$$
Here $\Omega^{MN}$ is the contravariant metric and
the second condition is the $q$-antisymmetry
property of the covariant symplectic metric. 
The action principle with the above action (2.21) leads to the 
 Hamilton equations:
$$
\begin{array}{lcl}
\dot z^M = \;    
{\displaystyle \sum_{N}} \Omega^{MN}\; \frac { \partial H} {\partial z^N}
\equiv \{ z^M, H \},
\end{array}\eqno(2.23)
$$
where the general $qp$-Poisson bracket is defined as [21,26]
$$
\begin{array}{lcl}
\{  F, G \} = 
{\displaystyle \sum_{M,N}} \Omega^{MN}\;\partial_{N}\;G~ \partial_{M}\;F,
\end{array}\eqno(2.24)
$$
with $ \partial_{M}= \frac{\partial}{\partial z^M} $.
A general form of the symplectic metric that satisfies (2.22) is
$$
\begin{array}{lcl}
\Omega^{MN} =
\left (\begin{array}{cc}
0, & (hq)^{-1} \\
-h^{-1}, & 0 \\   \end{array} \right),
\end{array}
$$
$$
\begin{array}{lcl}
\Omega_{MN} =
\left (\begin{array}{cc}
0, & - h \\
hq, & 0 \\   \end{array} \right) ,
\end{array}\eqno(2.25)
$$
where $h$ is an arbitrary function of $q$ and $p$. 
The dynamics formulated in equations (2.12)--(2.16) is obtained
if we choose $ h = p $.\\

\noindent
{\it 2.3 Examples}\\

\noindent
We  begin  with the  
general form for the Lagrangian as a polynomial of the second degree 
in velocity $(\dot x)$ and position $(x)$, namely;
$$
\begin{array}{lcl} 
L(x,\dot x) = {\displaystyle \sum_{n,m \geq 0}^{2}}
  \;a_{nm}\;\dot x^n\;x^m, \end{array}\eqno(2.26)
$$   
where $ a_{nm} $ are the time-independent noncommuting  parameters.
Since the terms $a_{00}$, $a_{10} x$ and $a_{11} \dot x x $ do not
appear in the equation of motion, we shall put 
$ a_{00} = a_{10} = a_{11} = 0$. The restrictions on the Lagrangian
given by (2.20) lead to rather complicated restrictions on 
 $a_{nm}$, in general. We first consider the simple case
when $ pq = 1$. Then, all the 
restrictions (2.20) are satisfied if we require the
following $q$-commutation relations between $a_{nm}$ and 
$\xi$ ($ = x, \dot x, \delta x, \delta \dot x$)
$$
\begin{array}{lcl}
\xi\;   a_{nm}  = q\; a_{nm}\; \xi, 
\end{array}\eqno(2.27)
$$ 
 and following amongst $a_{nm}$:
$$
\begin{array}{lcl}
 a_{02} a_{20} &=&  a_{20} a_{02}, \quad
 a_{12} a_{21}   =   q\;a_{21} a_{12},
\quad a_{01} a_{02} = q\; a_{02} a_{01},
\quad  a_{01} a_{20} =  q \; a_{20} a_{01}, \nonumber\\  
a_{01} a_{12} &=& q^2\;a_{12} a_{01},\quad
a_{01} a_{21}  =  q^2\; a_{21} a_{01}, \quad
a_{01} a_{22} = q^3\; a_{22} a_{01} , \quad
a_{02} a_{12} = q\; a_{12} a_{02}, \nonumber\\
a_{02} a_{21} &=& q\; a_{21} a_{02}, \qquad\
a_{20} a_{12} = q\; a_{12} a_{20}, \qquad
a_{20} a_{21}  =  q\; a_{21} a_{20}, \nonumber\\
a_{20} a_{22} & =& q^2 \; a_{22} a_{20}, \qquad
a_{12} a_{22} = q\; a_{22} a_{12}, \qquad
a_{21} a_{22} = q\; a_{22} a_{21}.
\end{array}\eqno(2.28)
$$
We now find the dynamical equation of motion
$$
\begin{array}{lcl}
( a_{20} + a_{21}\;x + a_{22}\;x^2 ) \; \ddot x
+ \frac{1}{2}\; ( a_{21} + 2\;a_{22}\;x )\;\dot x^2 - a_{02}\;x
- \frac{1}{2} a_{01} = 0.
\end{array} \eqno(2.29)
$$
 The cononical momentum  $ \pi  $ and the Hamiltonian $H$ are
$$
\begin{array}{lcl}
\pi  = q\;\bigl ( \frac{\partial L}{\partial \dot x } \bigr )
\equiv 2\;( a_{20} + a_{21}\; x + a_{22} x^2 )\;\dot x + a_{12}\;x^2,
\end{array} \eqno(2.30)
$$
$$
\begin{array}{lcl}
H = \frac{1}{4} ( \pi - a_{12} x^2 )\;
( a_{20} + a_{21}\;x + a_{22}\;x^2 )^{-1}
( \pi - a_{12} x^2 ) - a_{01}\; x - a_{02} x^2.
\end{array}\eqno(2.31)
$$
The Hamilton equations of motion are:
$$
\begin{array}{lcl}
\dot \pi = - q\; \frac{\partial H}{\partial x} 
= a_{01} + 2 a_{02} x + 2 a_{12} \dot x x 
+ 2 a_{22} \dot x^2 x + a_{21} \dot x^2, 
\end{array} \eqno(2.32)
$$
$$
\begin{array}{lcl}
\dot x = \frac{\partial H}{\partial \pi}
\equiv \frac{1}{2} ( a_{20} + a_{21}\;x + a_{22}\;x^2 )^{-1} 
( \pi - a_{12} x^2 ), 
\end{array}\eqno(2.33)
$$ 
and these are consistent with (2.29).\\

For further analysis, we specialise to the class of systems with
$ a_{12} = a_{21}
= a_{22} = 0$. The Lagrangian, the Hamiltonian, the canonical
momentum and the equation of motion are
$$
\begin{array}{lcl}
L  = a_{20}\; \dot x^2 + a_{02}\; x^2 + a_{01}\; x,
\end{array} \eqno(2.34)
$$
$$
\begin{array}{lcl}
H  = \frac{1}{4}\; \pi \; a_{20}^{-1}\; \pi
 - a_{01}\; x - a_{02}\; x^2, 
\end{array} \eqno(2.35)
$$
$$
\begin{array}{lcl}
\pi  = 2\; a_{20}\; \dot x, 
\end{array} \eqno(2.36)
$$
$$
\begin{array}{lcl}
 \ddot x  = 
a_{20}^{-1}\; a_{02}\;x
+ \frac{1}{2} a_{20}^{-1} a_{01} \equiv \omega^2 x + C,
\end{array}\eqno(2.37)
$$
where 
$$ 
\begin{array}{lcl}
\omega^2 = a_{20}^{-1} a_{02}
\end{array} \eqno(2.38)
$$ 
$$ 
\begin{array}{lcl}
C = \frac{1}{2} a_{20}^{-1}  a_{01}.
\end{array}\eqno(2.39)
$$
The solution is
$$
\begin{array}{lcl}
  x(t) = \; e^{\omega t}\; A  +  e^{-\omega t}\; B 
- \omega^{-2}\; C , 
\end{array}\eqno(2.40)   
$$ 
where $A$ and $B$ are constants which may be noncommuting in general and
they can be determined in terms of $x(0)$ and $\dot x(0)$:
$$
\begin{array}{lcl}
  A = \frac{1}{2}\;
\Bigl [\;  x(0) + \omega^{-2}\; C  
+ \omega^{-1}\; \dot x(0)\; \Bigr ], 
\end{array} \eqno(2.41)
$$
$$
\begin{array}{lcl}
 B = \frac{1}{2}\;
\Bigl [\;  x(0)  +  \omega^{-2}\; C - \omega^{-1}\; \dot x(0)\; \Bigr ].  
\end{array}\eqno(2.42)   
$$
Note that (2.38) defines only $\omega^2$, but we assume that 
$\omega$ and $\omega^{-1}$ also exist. The basic $q$-commutation
relation from (2.6) at $ t = 0$ is
$$
\begin{array}{lcl}
x(0)\; \dot x(0) = \dot x(0)\; x(0).
\end{array} \eqno(2.43)
$$ 
Using (2.43), (2.27) and (2.28), we find
$$
\begin{array}{lcl}
A\;B  = B\; A, \quad  B\;C = C\;B, \quad C\;A = A\;C,
\end{array} \eqno(2.44)
$$
$$
\begin{array}{lcl}
\omega\; A  = A\; \omega, \quad \omega\;B = B\; \omega,
\quad \omega\;C = C\; \omega,
\end{array}\eqno(2.45)
$$
$$
\begin{array}{lcl}
\omega\;x(0) = x(0) \;\omega, \quad 
\omega\; \dot x(0) = \dot x(0) \;\omega,
\end{array} \eqno(2.46)
$$
Further, it can be verified that all the $q$-commutation relations
among $x, \pi, \dot x $ and $\dot \pi$ are satisfied for all values
of $t$. {\it Thus, we have a completely consistent dynamics in a
noncommutative phase space, for the system defined by 
either (2.34) or (2.35).} However,
this consistency has been obtained at a price; all the constants
$ A, B, C $ and $ \omega $ that determine the solution $ x(t) $ in
(2.40), commute with each other.\\

Let us again consider the Lagrangian of (2.34) and ask whether
a more general dynamics with $pq \neq 1 $ can be constructed from it. We 
find that the restrictions (2.20) can be satisfied for $ p q \neq 1$ if 
we take the $q$-commutation relations of $a_{nm}$ with $\xi ( = x, \dot 
x, \delta x, \delta \dot x )$: $$
\begin{array}{lcl}
\xi\; a_{01} = q\; a_{01}\;\xi,
\quad \xi\; a_{02} = p\;q^2\; a_{02}\;\xi, \quad
\xi\; a_{20} = p^{-1}\; a_{20}\;\xi,
\end{array} \eqno(2.47)
$$
and $q$-commutation relations among $a_{nm}$
$$
\begin{array}{lcl}
a_{20}\;a_{01} = p\; a_{01}\; a_{20}, \qquad
a_{20}\;a_{02} = a_{02}\;a_{20},
\end{array} \eqno(2.48)
$$
but, we also require
$$
\begin{array}{lcl}
(pq - 1)\; ( a_{02}\; x^2 - pq\; a_{20}\; \dot x^2) = 0,
\end{array} \eqno(2.49)
$$
which follows from the relation
$$
\begin{array}{lcl}
x\;\dot \pi = q\; \dot \pi\; x + (pq -1)\;\dot x\;\pi.
\end{array}
$$
As a consequence of the 
restriction (2.49), dynamics for $ pq \neq  1$ does not evolve in the two-
dimensional phase space, but appears to degenerate into a restricted
one-dimensional region defined by (2.49).\\

Further study of the time evolution of the system confirms the above
conclusion. The equation of motion
$$
\begin{array}{lcl}
 \ddot x = 
 \frac{1}{p^2 q^2}\;\; a_{20}^{-1}\; a_{02}\;x
+\; \frac{1}{pq (1 + pq)}\;\; a_{20}^{-1} a_{01},
\end{array}\eqno(2.50)
$$
has the solution:
$$
\begin{array}{lcl}
x(t) = e^{\omega t}\; A + e^{-\omega t}\; B
- \omega^{-2}\; C,
\end{array} \eqno(2.51)
$$
where 
$$
\begin{array}{lcl}
\omega^2  = (pq)^{-2}\; a_{20}^{-1}\; a_{02},
\end{array}\eqno(2.52)
$$
$$
\begin{array}{lcl}
C = (pq)^{-1} (1 + pq)^{-1}\; a_{20}^{-1}\;a_{01},
\end{array} \eqno(2.53)
$$
and $A$ and $B$ can be again written as:
$$
\begin{array}{lcl} 
A = \frac{1}{2} \Bigl [ \; x(0) + \omega^{-2}\; C 
+ \omega^{-1} \;\dot x(0)\;
\Bigr ],
\end{array}\eqno(2.54) 
$$
$$
\begin{array}{lcl}
B = \frac{1}{2} \Bigl [\;
x(0) + \omega^{-2}\;C - \omega^{-1}\;
\dot x(0) \; \Bigr ].
 \end{array}\eqno(2.55)
$$
To satisfy (2.49) at $t=0$, we put
%$$
%\begin{array}{lcl}
%\dot x(0) = \pm\; \omega\; x(0).
%\end{array}\eqno(2.56)
%$$
$ C = 0 $ and so we drop the $a_{01}$ term in the
Lagrangian but, in addition, we require either $B = 0$ or $A = 0$.
As a consequence, we have

$$
\begin{array}{lcl}
\dot x(0) = \pm\; \omega\; x(0).
\end{array}\eqno(2.56)
$$

\noindent The corresponding solutions are therefore either
$$
\begin{array}{lcl}
x(t) = e^{\omega t}\; A  \quad  \mbox{with}
\quad A\;\omega = \;pq\; \omega\; A,
\end{array} \eqno(2.57)
$$
or,
$$
\begin{array}{lcl}
x(t) = e^{- \omega t}\; B \quad \mbox{with} \quad
B\;\omega = \;pq\; \omega\;B.
\end{array} \eqno(2.58)
$$
 One can verify that either of these solutions
satisfies all the consistency conditions including the
restriction (2.49) ($ a_{02} x^2 = pq \; a_{20} \dot x^2 $)
 for all $t$. Because of (2.56), these solutions do not allow
arbitrary and independent initial values for $ x(0)$ and
$ \dot x(0)$. In other words, the phase space point at $ t = 0$
 in the two-dimensional phase space must be chosen to 
lie in the restricted
``one dimensional phase space'' and the subsequent dynamics evolves 
within this one-dimensional space. Note that in contrast to the
case of $pq = 1$, the two constants $\omega$ and $A$ 
( or $B$) on which $x(t)$ depends, do not commute with each-other,
still there exists a consistent time evolution, albeit a 
restricted one.\\

Before we close this subsection, we remark that the examples discussed
here, not only generalize the ``q-deformed free particle  and
harmonic oscillator systems'' studied in Refs. [20] and [18],
but also incorporate them in a well-defined framework of
$q$-deformed differential calculus. Thus, our formalism
provides a sound basis for the heuristic results derived
earlier in the literature.\\

\noindent
{\bf 3 Multidimensional systems}\\

\noindent
{\it 3.1 Differential  calculus}\\

\noindent
We begin with an $N$ dimensional configuration space which is undeformed
( i.e., $ x_{i}\;x_{j} = x_{j}\;x_{i}, i,j = 1, 2, 3,...........N $). In the
corresponding $2N$ dimensional momentum phase space ( cotangent manifold ),
we introduce the $q$-deformation in such a way as to preserve the $GL(N)$
invariance in the configuration as well as the momentum space separately.
We take for all $i$ and $j$
$$
\begin{array}{lcl}
x_{i}\;x_{j}=x_{j}\;x_{i}, \quad  \pi_{i}\;\pi_{j}=\pi_{j}\;\pi_{i},
\quad x_{i}\;\pi_{j}= q\;\pi_{j}\;x_{i},
\end{array}\eqno(3.1)
$$
where $ \pi_{i} $ are the conjugate momenta corresponding to the coordinates
$x_{i}$  of the configuration space and $q$  is a nonzero  c-number. The
above set of relations are invariant under the  $GL(N)$
transformations
$$
\begin{array}{lcl}
x_{i}^{\prime} = a_{ij}\; x_{j}, \quad  \pi_{i}^{\prime} =  a_{ij}\; \pi_{j},
\end{array} \eqno(3.2)
$$
where $a_{ij}$ are the commuting c-number elements of a 
$ N \times  N $ nonsingular matrix of the undeformed general
 linear group of transformations $GL(N)$.  
 If we now apply the quantum group transformations
$$
\begin{array}{lcl}
x_{i} &\rightarrow& A\; x_{i} + B\; \pi_{i}, \nonumber\\
\pi_{i} &\rightarrow& C\; x_{i} + D\; \pi_{i}, \label{3.3}
\end{array}\eqno(3.3)
$$
where $A,B,C,D$ are the elements of a $ 2 \times 2 $ $GL_{q,p}(2)$ matrix
obeying the relations given in (2.3), the $q$-algebraic relations  (3.1)
remain form invariant only if $ pq = 1 $. Here the commutativity of the
group elements with the dynamical variables is assumed and the
transformations (3.3) are applied between pairs of conjugate variables:
$ (x_{1},\pi_{1}), (x_{2}, \pi_{2})....(x_{N},\pi_{N}) $. Under
the restriction $ pq = 1 $, the relations (2.3) reduce to a simpler form:
$$
\begin{array}{lcl}
A\;B &=& q^{-1}\; B\;A, \quad A\;C = q\; C\;A,
\quad B\;C = q^2\;C\;B,  \nonumber\\
C\;D &=&q^{-1}\;D\;C, \quad AD\;=\;DA, \quad B\;D = q\;D\;B.
\end{array}\eqno(3.4)
$$

We now develop the  differential calculus that is invariant under
(3.2) and (3.3). The differential calculus is based on the following 
 relations  in addition to (3.1):
$$
\begin{array}{lcl}
x_{i}\;\dot x_{j} &= & \dot x_{j}\;x_{i},
\quad  \pi_{i}\;\dot \pi_{j}=\dot \pi_{j}\;\pi_{i},
\quad x_{i}\;\dot \pi_{j}= q\;\dot \pi_{j}\;x_{i},
\quad  \pi_{i}\;\dot x_{j}=q^{-1}\;\dot x_{j}\;\pi_{i},\nonumber\\
x_{i}\;\delta x_{j}& = & \delta x_{j}\;x_{i},
\quad  \pi_{i}\;\delta \pi_{j}=\delta \pi_{j}\;\pi_{i},
\quad x_{i}\;\delta \pi_{j}= q\;\delta \pi_{j}\;x_{i},
\quad  \pi_{i}\;\delta x_{j}=q^{-1}\;\delta x_{j}\;\pi_{i},\nonumber\\
\dot x_{i}\;\delta x_{j}&=&\delta x_{j}\;\dot x_{i},
\quad \dot \pi_{i}\;\delta \pi_{j}=\delta \pi_{j}\;\dot \pi_{i},
\quad x_{i}\;\delta \dot x_{j}=\delta \dot x_{j}\; x_{i},
\quad  \pi_{i}\;\delta \dot \pi_{j}=\delta \dot \pi_{j}\; \pi_{i},
\nonumber\\
\dot x_{i}\;\delta \pi_{j}&=& q\;\delta \pi_{j}\;\dot x_{i},
\quad  \dot \pi_{i}\;\delta x_{j}=q^{-1}\;\delta x_{j}\;\dot \pi_{i},
\quad x_{i}\;\delta \dot \pi_{j} = q\;\delta \dot \pi_{j}\; x_{i},
\quad  \pi_{i}\;\delta \dot x_{j}=q^{-1}\;\delta \dot x_{j}\; \pi_{i},
\nonumber\\
\dot x_{i}\;\dot x_{j}&=&\dot x_{j}\;\dot x_{i},
\quad \dot \pi_{i}\;\dot \pi_{j}=\dot \pi_{j}\;\dot \pi_{i},
\quad \dot x_{i}\;\delta \dot x_{j}=\delta \dot x_{j}\; \dot x_{i},
\quad \dot \pi_{i}\;\delta \dot \pi_{j}=\delta \dot \pi_{j}\; \dot \pi_{i},
\nonumber\\
\dot x_{i}\;\dot \pi_{j}&=& q\;\dot \pi_{j}\;\dot x_{i},
\quad  \dot \pi_{i}\;\dot x_{j}=q^{-1}\;\dot x_{j}\;\dot \pi_{i},
\quad \dot x_{i}\;\delta \dot \pi_{j} = q\;\delta \dot \pi_{j}\; \dot x_{i},
\quad  \dot \pi_{i}\;\delta \dot x_{j}
=q^{-1}\;\delta \dot x_{j}\; \dot \pi_{i}.
\end{array}\eqno(3.5)
$$
The following wedge products complete the whole algebra:
$$
\begin{array}{lcl}
\delta x_{i}\;\wedge\;\delta x_{j}
&=& -\; \delta  x_{j}\; \wedge\;\delta x_{i},\quad
\delta \pi_{i}\;\wedge\;\delta \pi_{j}
= -\; \delta  \pi_{j}\;\wedge\; \delta \pi_{i}, \nonumber\\
\delta \dot x_{i}\;\wedge\;\delta \dot x_{j}
&=& -\; \delta  \dot x_{j}\; \wedge\;\delta \dot x_{i},\quad
\delta \dot \pi_{i}\;\wedge\;\delta \dot \pi_{j}
= -\; \delta \dot \pi_{j}\;\wedge\; \delta \dot \pi_{i}, \nonumber\\
\delta  x_{i}\;\wedge\;\delta \dot x_{j}
&=& -\; \delta  \dot x_{j}\; \wedge\;\delta  x_{i},\quad
\delta  \pi_{i}\;\wedge\;\delta \dot \pi_{j}
= -\; \delta \dot \pi_{j}\;\wedge\; \delta  \pi_{i}, \nonumber\\
\delta \pi_{i}\;\wedge\;\delta \dot \pi_{j}
&=& -\; \delta \dot \pi_{j}\;\wedge\; \delta \pi_{i}, \quad
\delta x_{i}\;\wedge\;\delta \pi_{j}
= -q\; \delta \pi_{j}\; \wedge\;\delta x_{i},\nonumber\\
\delta \dot x_{i}\;\wedge\;\delta \dot \pi_{j}
&=& -q\; \delta \dot \pi_{j}\; \wedge\;\delta \dot x_{i},  \quad
\delta \pi_{i}\;\wedge\;\delta \dot x_{j}
= -\; q^{-1}\;\delta \dot x_{j}\;\wedge\; \delta \pi_{i}, \\
 \delta x_{i}\;\wedge\;\delta \dot \pi_{j} 
&=& -q\; \delta \dot \pi_{j} \;\wedge\;\delta x_{i}. \\
\end{array}\eqno(3.6)
$$
All the relations in (3.5) and (3.6) are
invariant under the $GL_{q,q^{-1}}$ transformations (3.3) applied on
the canonical pairs $ (x_{i}, \pi_{i}), (\delta x_{i}, \delta \pi_{i}),
(\dot x_{i}, \dot \pi_{i})$ or 
$ (\delta \dot x_{i}, \delta \dot \pi_{i})$ for each $i$ and are also
invariant  under the $GL(N)$ transformations defined by (3.2) together
with the following:  
$$
\begin{array}{lcl}
\delta x_{i}^{\prime} &=& a_{ij}\; \delta x_{j}, \quad
\delta \pi_{i}^{\prime} =  a_{ij}\; \delta \pi_{j}, \quad
\dot x_{i}^{\prime} = a_{ij}\; \dot x_{j}, \nonumber\\
\delta \dot x_{i}^{\prime} &=& a_{ij}\; \delta \dot x_{j},\quad
\delta \dot \pi_{i}^{\prime} =  a_{ij}\; \delta \dot \pi_{j}, \quad
\dot \pi_{i}^{\prime} = a_{ij}\; \dot \pi_{j}.
\end{array}\eqno(3.7)\\
$$

The above differential calculus has been constructed in such a way as to
preserve the separate $GL(N)$ invariance of the configuration space and
the momentum space. As a consequence, we will be able to construct
dynamics in which rotational or Lorentz invariance could be 
preserved.\\ 

\noindent
{\it 3.2 Dynamics} \\

\noindent
Once again we start with the action
$$
\begin{array}{lcl}
S = \int dt\; \Bigl [\; {\displaystyle \sum_{i}\;}
\pi_{i}\;\dot x_{i} - H (x_{i},\pi_{i})
\; \Bigr ].
\end{array}\eqno(3.8)
$$
Using the multidimensional generalizations  
of (2.4) and (2.5), the action principle
$$
\begin{array}{lcl}
\delta S = 0 = \int dt\; \Bigl [ \; {\displaystyle \sum_{i}}
\bigl (\; \delta \pi_{i}\; \dot x_{i} \;+\; \pi_{i}\; \delta \dot x_{i} \;
- 
\; \delta x_{i}\; {\displaystyle \frac{ \partial H}{ \partial x_{i}}}
-\; \delta \pi_{i}\; {\displaystyle \frac{\partial H}{\partial \pi_{i}}} \;
\bigr ) \; \Bigr ],
\end{array}\eqno(3.9)
$$
leads to the following Hamilton equations of motion
$$
\begin{array}{lcl}
\dot x_{i}  =  \; { \displaystyle \frac{\partial H}
{\partial \pi_{i}}}, 
\end{array}
$$
$$
\begin{array}{lcl}
\dot \pi_{i} &=& -\;q\;
 {\displaystyle \frac{\partial H}{\partial x_{i}}},
\end{array} \eqno(3.10)
$$
if we exploit the $q$-algebraic relation
$$ 
\begin{array}{lcl}
 \pi_{i}\;\delta \dot x_{i}=
 -\;q^{-1} \delta \dot x_{i}\;\pi_{i}, 
\end{array}  
$$ 
{}from (3.5). The Lagrangian function $ L(x_{i},\dot x_{i} )$ can be 
defined as $$
\begin{array}{lcl}
 L(x_{i},\dot x_{i}) = {\displaystyle \sum_{i}}\;\pi_{i}\;\dot x_{i}
- H (x_{i},\pi_{i}), 
\end{array}\eqno(3.11)
$$
where $\pi_{i}$ must be expressed in terms of $x_{i}$ 
and $\dot x_{i}$ through
$$
\begin{array}{lcl}
\pi_{i} = q\; \Bigl ( 
{\displaystyle \frac{\partial L} { \partial \dot x_{i}}} \Bigr ).\\
\end{array} \eqno(3.12)
$$

 The general form of the $qp$-Poisson
bracket for the dynamical variables $ F(x_{i},\pi_{i}) $ and 
$ G(x_{i},\pi_{i}) $
is
$$
\begin{array}{lcl}
\{ F, G \}  = {\displaystyle \sum_{i}} \Bigl (  
\frac{\displaystyle \partial G}
{\displaystyle \partial \pi_{i}}   
 \frac{\displaystyle 
\partial F} {\displaystyle \partial x_{i}} -
\;q\; 
\frac{\displaystyle \partial G}{\displaystyle \partial x_{i}} 
 \frac{\displaystyle \partial F}
{\displaystyle \partial \pi_{i}} \Bigr ).
\end{array}\eqno(3.13)
$$
and (3.10) can be rewritten as
$$
\begin{array}{lcl}
\dot x_{i} &=& \{ x_{i}, H \}, \nonumber\\
\dot \pi_{i} &=& \{ \pi_{i}, H \}.
\end{array}\eqno(3.14)
$$

\noindent The brackets between canonical pairs are:
$$
\begin{array}{lcl}
\{ x_{i}, \pi_{j} \} &=&  \; \delta_{ij},
 \nonumber\\
\{ \pi_{i}, x_{j} \} &=& -\;q \; \delta_{ij},
\nonumber\\
\{ x_{i}, x_{j} \} &=& \{ \pi_{i}, \pi_{j} \} = 0,
\end{array}\eqno(3.15)
$$
%and (3.9) can be rewritten as
%$$
%\begin{array}{lcl}
%\dot x_{i} &=& \{ x_{i}, H \}, \nonumber\\
%\dot \pi_{i} &=& \{ \pi_{i}, H \}.
%\end{array}\eqno(3.15)
%$$

The rest of the  discussion proceeds on similar lines as discussed in
Sec. 2 for the one-dimensional case. The following six
 relations from (3.5)
$$
\begin{array}{lcl}
\pi_{i}\;\pi_{j} &=& \pi_{j}\;\pi_{i}, \;\quad \;
x_{i}\;\pi_{j} =q\; \pi_{j}\;x_{i}, \nonumber\\
\pi_{i}\;\dot \pi_{j}&=& \; \dot \pi_{j}\;\pi_{i},\;\quad \;
\pi_{i}\;\dot x_{j} = q^{-1}\;\dot x_{j}\;\pi_{i}, \;\nonumber\\
x_{i}\;\;\dot \pi_{j} &=& q\; \dot \pi_{j}\; x_{i}
,\quad
\dot x_{i}\;\dot \pi_{j} =q\; \dot \pi_{j}\;\dot x_{i},
\end{array}
$$
lead to  the restrictions on the Lagrangian:
$$
\begin{array}{lcl}
\frac{\displaystyle \partial L}{\displaystyle \partial \dot x_{i}}\;
\frac{\displaystyle \partial L}{\displaystyle \partial \dot x_{j}} =
\frac{\displaystyle \partial L}{\displaystyle \partial \dot x_{j}}
\frac{\displaystyle \partial L}{\displaystyle \partial \dot x_{i}}\quad
x_{i}\;\frac{\displaystyle \partial L}{\displaystyle \partial \dot x_{j}}
 =q\; \frac{\displaystyle \partial L}{\displaystyle \partial \dot x_{j}}
\;x_{i}, 
\end{array}
$$
$$
\begin{array}{lcl}
\frac{\displaystyle \partial L}{\displaystyle \partial \dot x_{i}}\;
\frac{\displaystyle \partial L}{\displaystyle \partial  x_{j}} =\;
\frac{\displaystyle \partial L}{\displaystyle \partial  x_{j}}
\frac{\displaystyle \partial L}{\displaystyle \partial \dot x_{i}}, \qquad
\frac{\displaystyle \partial L}{\displaystyle \partial \dot x_{i}}\;\dot 
x_{j} = q^{-1}\;\dot x_{j}\; 
\frac{\displaystyle \partial L}{\displaystyle \partial \dot x_{i}},
\end{array}
$$
$$
\begin{array}{lcl}
x_{i}\;\;\frac{\displaystyle \partial L}
{\displaystyle \partial  x_{j}} = q\;
\frac{\displaystyle \partial L}{\displaystyle \partial  x_{j}}\; x_{i},
\quad
\dot x_{i}\;\frac{\displaystyle \partial L}
{\displaystyle \partial  x_{j}} = q\;
\frac{\displaystyle \partial L}
{\displaystyle \partial  x_{j}}\;\dot x_{i}.
\end{array}\eqno(3.16)\\
$$

The above dynamics can be reexpressed in
terms of the  $q$-deformed symplectic
manifold in the multidimensional phase space. This is defined through  
the symplectic metric
$$
\begin{array}{lcl}
\Omega^{MN} =
\left (\begin{array}{cc}
0, & 1 \\
-q, & 0 \\   \end{array} \right) ,\quad
\end{array}
$$
$$
\begin{array}{lcl}
\Omega_{MN} =
\left (\begin{array}{cc}
0, & -q^{-1} \\
1, & 0 \\   \end{array} \right) ,
\end{array}\eqno(3.17)
$$
associated with each canonical pair: 
$(x_{1},\pi_{1}), (x_{2}, \pi_{2})...........(x_{N},\pi_{N})$.
Thus, there are $N$ copies of the above metric for the whole
$2N$ dimensional cotangent manifold. 
In terms of the above metric, the general form of the 
Poisson bracket and Legendre transformation are
$$
\begin{array}{lcl} 
\{  F, G \} = {\displaystyle \sum_{i,M,N} }
\Omega^{MN}\;
\partial_{Ni}\; G~ 
\partial_{Mi}\;F ,
\end{array} \eqno(3.18)
$$
$$
\begin{array}{lcl}
L(x_{i},\dot x_{i})& =& \frac{1} {2}\;\;{\displaystyle \sum_{i,M,N}} 
\Omega_{MN}\;z^{M}_{i}\;\dot z^{N}_{i}
-\;   H(z_{i}),\nonumber\\
&=& 
{\displaystyle \sum_{i}}\; \Bigl ( \frac{1}{2} \pi_{i}\;\dot x_{i}
- \frac{1} {2q}\; x_{i}\; \dot \pi_{i} \Bigr )  - H (x_{i},\pi_{i}),
\end{array}\eqno(3.19)
$$
with $ \partial_{Mi}= \frac{\partial}{\partial z^M_{i}} $ and 
$ z^M_{i} = (x_{i},\pi_{i}) $.
The Hamilton equations can be derived from the least action principle
and can be expressed concisely as:
$$
\begin{array}{lcl}
\dot z^{M}_{i} =  {\displaystyle \sum_{N}}
\Omega^{MN}\; \frac{\partial H} {\partial z^{N}_{i}}
= \{ z_{i}^M , H \}.
\end{array}\eqno(3.20)\\
$$

\noindent
{\it 3.3 Examples}\\

\noindent
For the multidimensional case, right from the beginning, we have the 
restriction $pq=1$ since this is required by the  invariance of (3.1)
under the quantum group . Thus, the example of a 
consistent one-dimensional system given by (2.34) or (2.35) 
can be readily generalized to the multidimensional case. Let us
take 
$$
\begin{array}{lcl}
L = a_{20}\;{\displaystyle \sum_{i}} \dot x_{i}\; \dot x_{i}
  + a_{02}\;{\displaystyle \sum_{i}} x_{i}\;x_{i}.
\end{array} \eqno(3.21)
$$
All the $q$-commutation relations
remain intact in the multidimensional case except that
the one dimensional canonical pair ( $x,\pi$ ) has to be replaced by the
multidimensional$~~$ pair: ( $ x_{i}, \pi_{i} $ ). The solution
$$
\begin{array}{lcl}
x_{i}(t) = e^{\omega t}\; A_{i} + e^{- \omega t}\; B_{i},
\end{array} \eqno(3.22)
$$
with $ \omega, A_{i}$ and $ B_{i} $
commuting with each other, gives a completely consistent dynamical
evolution. This dynamics, although $q$-deformed, is invariant under
the undeformed $SO(N)$ rotations since the Lagrangian in (3.21) is $SO(N)$
invariant.\\

Finally we  discuss the $q$-deformed free massless relativistic particle
in $D$-dimensional space-time as another example of multidimensional 
systems. 
The trajectory of the  particle
is parameterized by a commuting ``proper time'' evolution parameter
$\tau$ and it is embedded in a flat Minkowski 
$D$-dimensional space-time target manifold.
The corresponding $q$-deformed cotangent manifold  is characterized
by $GL_{q,q^{-1}}(2)$ invariant relations ( same as  (3.1)):
$$
\begin{array}{lcl}
x_{i} x_{j} = x_{j} x_{i}, \quad 
\pi_{i} \pi_{j} = \pi_{j} \pi_{i}, \quad
x_{i} \pi_{j} = q\; \pi_{j} x_{i}.
\end{array}
$$
The  Lagrangian and Hamiltonian for
this system are [20]
$$
\begin{array}{lcl}
L = \frac{1}{2}\;q^{-1}\;e^{-1}\;\dot x^2, \quad
H = \frac{1}{2}\;e\;\pi^2,
\end{array}\eqno(3.23)
$$
where $e$ is an einbein field and $\dot x_{i}$ and $\pi_{i}$ are
velocities and momenta of the particle in the $D$-dimensional 
%target space with $ \dot x_{i} = 
target space with  

$$
\begin{array}{lcl}
\dot x_{i} = \frac{\displaystyle \partial x_{i}} {\displaystyle \partial \tau}
\end{array} \eqno(3.24)
$$

$$
\begin{array}{lcl}
\dot x^{2} \equiv \sum^{D-1}_{i=1} \dot x_{i} \dot x_{i} 
- \dot x_{D} \dot x_{D} ; \quad 
\pi^2 \equiv \sum^{D-1}_{i = 1} \pi_{i} \pi_{i} - \pi_{D} \pi_{D}
\end{array} \eqno(3.25)
$$

\noindent To prove the consistency
of the above expressions, we exploit the following relations from the
differential calculus ( Subsec. 3.1)
$$
\begin{array}{lcl}
\dot x_{i} \; x_{j} = x_{j}\; \dot x_{i},\quad 
\pi_{i}\;\pi_{j} = \pi_{j}\;\pi_{i}, \quad 
\dot x_{i}\; \dot x_{j} = \dot x_{j}\; \dot x_{i},\quad
\dot x_{i}\; \pi_{j} = q \;\pi_{j}\;\dot x_{i}, \label{3.21}
\end{array}\eqno(3.26)
$$
which lead to the following set of $q$-algebraic relations
 among einbein,
coordinates, momenta and velocities:
$$\begin{array}{lcl}
e \dot x_{i}& =& q\; \dot x_{i} e, \quad  e \;x_{i} = q \; x_{i}\; e,
\quad \delta e\; \dot x_{i}= q \dot x_{i}\; \delta e, \quad
\delta e \;\pi_{i} = q \pi_{i}\; \delta e,\nonumber\\
e \; \pi_{i}& = &q\; \pi_{i}\; e, \qquad\; e\;\delta \pi_{i} 
= q \delta \pi_{i} \; e, \qquad\; e\;\delta \dot x_{i} 
= q \; \delta \dot x_{i}\; e, 
\end{array}\eqno(3.27)
$$ 
if we exploit the equations of motion ( on-shell conditions )
$$
\begin{array}{lcl}
\dot x_{i} = q\;e\;\pi_{i}, \quad \dot \pi_{i} = 0,
\quad \dot x^2 = 0
\end{array} \eqno(3.28)
$$
Further, we have used
$$
\begin{array}{lcl}
\pi_{i} = q\;\Bigl ( 
{\displaystyle \frac{ \partial L} { \partial \dot x_{i}}} \Bigr ),
\quad \dot x_{i} = 
{\displaystyle \frac{ \partial H} {\partial \pi_{i}} }.
\end{array} 
$$
It can be verified that the Legendre transformation
$ L(x_{i},\dot x_{i}) = \sum_{i} \; \pi_{i}\; x_{i} - H (x_{i},\pi_{i}) $
is consistent if we use the on-shell conditions (3.28). 
Thus, we see that  the $q$-deformed massless 
relativistic particle can be formulated within our framework
and its dynamics is invariant under the $ D $ - dimensional Lorentz
transformation $ SO(D-1, 1)$. \\

\noindent
{ \bf 4 Summary and Discussion }\\ 

\noindent
We have generalized one-dimensional and multidimensional
 dynamics to a noncommutative phase space.
For a system with  one-dimensional configuration space, we have taken 
the phase space to be invariant under the quantum group $GL_{q,p}(2)$
with two independent deformation parameters $q$ and $p$, whereas
for the multidimensional case the quantum group invariance requires
$ p = q^{-1}$. A general formulation, using a deformed symplectic
manifold, has been given for the dynamics. We have 
succeeded in constructing examples of completely consistent 
$q$-deformed dynamical systems.\\

Nevertheless, it is important to point out a drawback of all forms
of $q$-deformed dynamics, including the ones constructed in Refs.
[18--20]. This is the existence of the restrictions on the Lagrangians
( (2.20) or (3.16) ). It is these restrictions that prevent us from
constructing a genuine $q$-deformed dynamical system for $pq \neq 1$
in Sec. 2.3. It is desirable to remove all such restrictions from 
dynamics. Hopefully further work will achieve it.\\

We have argued elsewhere [27] that $q$-deformation of the algebra of
creation and annihilation operators ( or equivalently, $q$-deformation
of the Heisenberg algebra ) does not lead to anything fundamentally new 
since it merely amounts to a redefinition of the operators acting on
the same Fock space ( or Hilbert space ). However, this argument is not 
quite applicable to the present work which is based on  classical
dynamics $(\hbar = 0)$. \\

It is widely speculated that  space-time structure will be 
modified at the Planck scale.
Although this is sometimes considered as a motivation for quantum-group 
based work, 
it is not clear how $q$-
deformation with a dimensionless parameter $q$ can lead to
such a modification. Further, it can be argued [28] that,
rather than any $q$-deformed algebra of the type (3.1) with a
specific commuation relation for the pair $(x_{i}, x_{j})$,
an algebra  that does not specify the
commutation relation for the pair $(x_{i},x_{j})$ may be more
relevant at the Planck scale since it would give rise to a larger 
framework. 
For instance, Greenberg's algebra [29,27] of creation and annihilation
operators ($a_{i}^{\dagger}, a_{i})$ that specifies only the 
commutation relations between $a_{i}^{\dagger}$ and $a_{i}$ but
leaves the commutation relations among $a_{i}'s$ free and unspecified,
leads to an enlargement of the conventional Fock space.
In spite of these arguments, it
is worthwhile to see how far can one push the dynamics based on
the $q$-deformation. It is in this spirit of exploration that
the present attempt has been made.\\

In our investigation, we  started with classical dynamics in which
$x$ and $\pi$ are commuting c-numbers and then ``quantized'' the dynamics
through $q$-deformation which turns $x$ and $\pi$ into
noncommuting operators. On what do these operate? What is the physical
meaning of these operators? These are deep questions that are yet
to be answered.\\
 
\noindent
{\bf Acknowledgement}\\

\noindent
One of us (RPM) would like to express his deep sense of gratitude to
the Director of JINR for giving him permission 
and financial support to visit IMSc
and the Director of IMSc for 
warm hospitality at Madras where this work was completed.\\

\end{document}